\documentclass[prl,twocolumn,superscriptaddress]{revtex4}
\usepackage{amsmath,amssymb,mathrsfs}
\usepackage[dvips]{graphicx}
\usepackage{hyperref}
\usepackage{paralist}

\newcommand{\ua}{\uparrow}
\newcommand{\da}{\downarrow}

\def\be{\begin{equation}}
\def\ee{\end{equation}}
\def\bea{\begin{eqnarray}}
\def\eea{\end{eqnarray}}

\begin{document}

\title{Wilson ratio of Fermi gases in one dimension}

\author{X.-W. Guan}
\email[e-mail:]{xwe105@physics.anu.edu.au}
\affiliation{State Key Laboratory of Magnetic Resonance and Atomic and Molecular Physics, 
Wuhan Institute of Physics and Mathematics, Chinese Academy of Sciences, Wuhan 430071, China}
\affiliation{Department of Theoretical Physics, Research School of Physics and Engineering, 
Australian National University, Canberra ACT 0200, Australia}

\author{X.-G. Yin}
\affiliation{Division of Materials Science, Nanyang Technological University, Singapore 639798}

\author{A. Foerster}
\affiliation{Department of Theoretical Physics, Research School of Physics and Engineering, 
Australian National University, Canberra ACT 0200, Australia}
\affiliation{Instituto de Fisica da UFRGS, Av. Bento Goncalves 9500, Porto Alegre, RS, Brazil }

\author{M. T. Batchelor}
\affiliation{Centre for Modern Physics, Chongqing University, Chongqing 400044, China}
\affiliation{Department of Theoretical Physics,
Research School of Physics and Engineering,
Australian National University, Canberra ACT 0200, Australia}
\affiliation{Mathematical Sciences Institute, Australian
National University, Canberra ACT 0200, Australia}

\author{C.-H.  Lee}
\affiliation{State Key Laboratory of Optoelectronic Materials and Technologies, School of Physics and Engineering, 
Sun Yat-Sen University, Guangzhou 510275, China}

\author{H.-Q. Lin}
\email[e-mail:]{haiqing0@csrc.ac.cn}
\affiliation{Beijing Computational Science Research Center, Beijing 100084, China}

\date{\today}


\begin{abstract}
We calculate the Wilson ratio of the one-dimensional Fermi gas with spin imbalance. 
The  Wilson ratio of attractively interacting fermions  is solely determined by the density stiffness and sound velocity of pairs   
and of excess  fermions  for the two-component Tomonaga-Luttinger liquid (TLL) phase.  
The ratio exhibits anomalous enhancement  at  the two critical points due to the sudden change in the density of states.
Despite a  breakdown of the quasiparticle description  in one dimension,  two important features of  the Fermi liquid are retained, 
namely the specific heat is linearly proportional to temperature  whereas  the susceptibility is independent of temperature.    
In contrast to the phenomenological TLL parameter, the Wilson ratio provides a powerful parameter for testing  
universal quantum liquids of  interacting fermions in one, two and three dimensions.
\end{abstract}


\maketitle
Fermi liquid theory describes the low-energy physics of interacting fermions, conduction electrons, heavy fermion metals  
and liquid ${}^3$He \cite{Hewson1997}. 
It is remarkable that  the Wilson ratio, defined as the ratio of the magnetic susceptibility $\chi$  to specific heat $c_v$ divided by temperature $T$,
\begin{equation}
R_W=\frac{4}{3}\left(\frac{\pi k_B}{\mu_B g }\right)^2\frac{\chi}{c_v/T}
\label{ratio}
\end{equation}
is a constant at the renormalization  fixed point of these systems. 
Here $k_B$ is the Boltzmann constant, $\mu_B$ is the Bohr magneton and $g$ is the Lande factor.  
For example, $R_W=1$ for noninteracting or weakly correlated electrons in metals \cite{Hewson1997}, 
and $R_W=2$ in the Kondo regime for  the impurity problem  \cite{Wilson1975}.
The dimensionless Wilson ratio quantifies the interaction effect and spin
fluctuations and thus presents a characteristic of strongly correlated
Fermi liquids \cite{Hewson1997}.
$R_W > 1$ in strongly correlated systems where the spin fluctuations are
enhanced while charge fluctuations are suppressed.

The Wilson ratio has recently been measured in experiments on a gapped spin-1/2 Heisenberg ladder \cite{Ninios2012}.
This opens up the opportunity to probe and understand the universal nature of  one-dimensional (1D) quantum liquids through the measurable Wilson ratio.
Early calculations of $R_W$ for 1D correlated electrons were considered only in the scenario of spin-charge separation \cite{Usuki1989, Schulz1991}.
As far as the low energy physics is concerned, the fixed point critical Tomonaga-Luttinger liquid (TLL) behaves much like the Fermi liquid \cite{WangYP}. 
For instance, the Wilson ratio  of the quasi-1D  spin-1/2 Heisenberg ladder near the critical point indicates a single component TLL with $R_W=4K$, 
where $K$ is the TLL parameter.  
Moreover, the Wilson ratio is always less than 2 as the band fillings tend towards the Mott insulator in the 1D repulsive Hubbard model \cite{Usuki1989}. 
For the 1D spin-1/2 Heisenberg chain $R_W=2$ as  $T \to 0$ \cite{Johnston}. 
Here the Fermi liquid nature arises because the elementary excitations at
low temperatures are spinons which are regarded as fermions.

Motivated by the experimental results for the spin ladder \cite{Ninios2012}, we consider the Wilson ratio in the context of the 
the spin-$1/2$ delta-function interacting Fermi gas \cite{Gaudin,Yang}.
The quantum liquids exhibited by this model include the  paradigm of a spin-charge separated TLL in the repulsive regime and a two-component TLL 
of  pairs and single fermions in the attractive regime. 
The pairing phase has attracted a great deal of attention \cite{Feiguin,Rizzi2008,Zhao2008,Lee2011,Schlottmann2012,Bolech2012,Lu2012}, with the 
key features of the $T = 0$ pairing phase \cite{Orso,HuiHu,Guan2007} experimentally confirmed using finite temperature density profiles of 
trapped fermionic ${}^6$Li atoms \cite{Liao,GBL}.

\begin{figure}[htbp]
\includegraphics[width=1.0\linewidth]{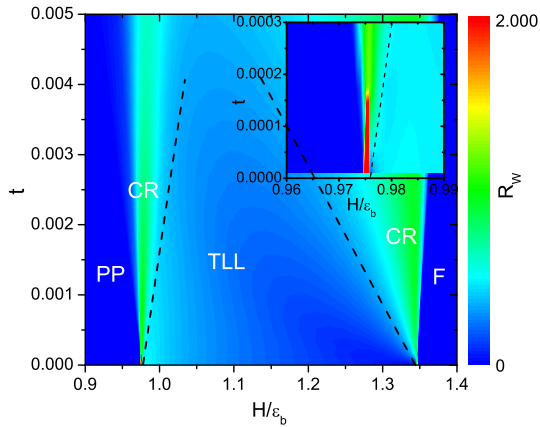}
\caption{(Color online) Contour plot of the Wilson ratio $R_W$ (\ref{ratio}) of the attractive Fermi gas for dimensionless interaction $|\gamma| =10$ 
as a function of the reduced temperature $t=T/\varepsilon_b$ and magnetic field.  $\varepsilon_b$ is the binding energy. 
The result (\ref{W_R}) provides a criterion for  the two-component TLL phase in the region below the dashed lines, 
where $R_W$ is temperature independent.  
The dashed lines indicate the crossover  temperature $T^*\sim |H-H_c|$ separating the relativistic liquid  from the nonrelativistic liquid. 
$R_W =0$ for both the TLL  of pairs (PP) and the TLL of excess fermions (F). In the critical regimes (CR) $R_W$ gives a 
temperature-dependent scaling. However, near the two critical points, the ratio reveals anomalous enhancement discussed further 
in the text.  The inset shows  the enhancement at  the lower critical point. } \label{fig:contour}
\end{figure}

In this context the Wilson ratio of the 1D attractive Fermi gas with polarization is particularly interesting due to the coexistence of pairing 
and depairing under the external magnetic field.    
It is natural to ask if the Wilson ratio can capture a similar Fermi liquid nature of such a particular pairing phase.  
Here we report  our key result  for the attractive Fermi gas,  
\begin{equation}
R_W=\frac{4}{\left( v_N^{\rm b}+4v_N^{\rm  u}\right)\left(\frac{1}{v_s^{\rm b}}+\frac{1}{v_s^{\rm  u}}\right)}\label{W_R}
\end{equation} 
which holds throughout the two-component TLL phase. 
This result is in terms of the density stiffness $v_N^{{\rm b,u}}$ and sound velocity $v_s^{{\rm b,u}}$ for 
pairs b and excess single fermions u. These parameters can be calculated from the ground state energy.  
Fig.~\ref{fig:contour} shows that at finite temperatures  the contour plot of $R_W$ can map out not only  the two-component TLL 
phase but also  the quantum criticality of the attractive Fermi gas.   
The Wilson ratio thus gives a simple testable parameter to quantify  interaction effects and the competing order between pairing and depairing.

\textit{The Model.-} The $\delta$-interacting  spin-1/2 Fermi gas with $N=N_{\uparrow}+N_{\downarrow}$ fermions of mass $m$ 
with external magnetic field $H$ is described by the Hamiltonian \cite{Gaudin,Yang,GBL}
\begin{eqnarray}
\mathcal{H} &=&-\frac{\hbar ^{2}}{2m}\sum_{i=1}^{N}\frac{\partial ^{2}}{
\partial x_{i}^{2}}+g_{1D}\sum_{i=1}^{N_{\uparrow}}\sum_{j=1}^{N_{\downarrow }}\delta \left( x_{i}-x_{j}\right)+ E_z \quad
\label{Hamiltonian}
\end{eqnarray}
in which the terms are the kinetic energy, interaction energy and Zeeman
energy $E_z= -\frac{1}{2}g \mu_BH\left( N_{\uparrow }-N_{\downarrow }\right)$. 
Here the inter-component interaction is determined by  an
effective 1D scattering length $g_{1D}=-\frac{2\hbar^2}{ma_{1D}}$ which can
be tuned from the weakly interacting regime ($g_{1D}\rightarrow 0^{\pm}$) to the
strong coupling regime ($g_{1D}\rightarrow \pm \infty $) via Feshbach
resonances and optical confinement \cite{Olshanii1998}. 
$g_{1D}>0$ ($<0$) is the contact repulsive (attractive) interaction.
The total density $n=n_\ua+n_\da$, the magnetization $M=(n_\ua-n_\da)/2$, 
and the polarization $P=(n_\ua-n_\da)/n$, where $n=N/L$ is the linear density and $L$  is the length of the system.
For convenience, we define the interaction strength as $c=mg_{1D}/\hbar ^{2}$ and dimensionless parameter 
$\gamma =c/n$ for physical analysis. 
We set Boltzmann constant $k_{B}=1$ and $\mu_B g=1$.

The thermodynamic properties of the model are determined by the thermodynamic 
Bethe ansatz (TBA) equations  \cite{Takahashi}.   
%
A high precision equation of state in the physically interesting low temperature and strong coupling regime 
($T\ll \epsilon _{b},H\text{ \ and \ }\gamma \gg 1$)  has been derived \cite{Zhao,Guan-Ho}.
The  hydrodynamic description of the attractive gas (\ref{Hamiltonian}) is restricted to the limit cases $c\to -\infty$ and $c\to 0^-$ \cite{Vekua}.

\textit{Susceptibility.-} In the Fermi liquid, the interaction enters the susceptibility and specific heat via the 
effective mass and the Landau parameters \cite{Schofield1999}. 
Thus the specific heat increases linearly with the temperature $T$ because  only the electrons  
within $k_BT$ near the Fermi surface contribute to the specific heat. 
The susceptibility is independent of temperature since only the electrons within $\mu_B g H$  
near the Fermi surface contribute to the magnetization.  
This is a consequence of the forward scattering process between quasiparticles near the Fermi surface.
In contrast, in 1D many-body systems, all particles participate in the low energy physics and thus form 
collective motion of bosons, i.e., the TLL.  
However, the TLL is also the consequence of the forward scattering process involving 
low-lying excitations close to Fermi points. 
Therefore it is natural to expect that 1D many-body systems have a Fermi liquid nature in the low energy sector.

%
%

Here we find such a Fermi liquid signature of the 1D Fermi gas using the analytic results for  the 
susceptibility and specific heat obtained via the TBA  equations \cite{ref-sup}. 
At zero temperature,  the susceptibility can be calculated  from the dressed energy equations
which are obtained from the TBA equations  in the
limit $T\to 0$ \cite{ref-sup}.  
The dressed energy  equations give the full phase diagram and magnetic properties in the grand  canonical ensemble.

For values of the magnetic field  between the  lower and upper  critical fields $H_{c1}$ and $H_{c2}$   
the zero temperature susceptibility of the gapless phase can be expressed in the form
 \begin{equation}
 \frac{1}{\chi}=\frac{1}{\chi_{\rm u}}+\frac{1}{\chi_{\rm b}}.\label{susceptibility}
 \end{equation}
This result can be established on general grounds.
The effective magnetic field $H$ depends on the chemical potential bias $H:=\Delta\mu =\mu_{\uparrow}-\mu_{\downarrow}$. 
The magnetization depends on the difference  $\Delta n=n_{\uparrow}-n_{\downarrow}$. 
We prove  that  the magnetic susceptibility $\chi=\frac{1}{2}\partial \Delta n/\partial \Delta \mu$ 
can be written in terms of the charge susceptibilities of bound pairs and excess fermions 
$\chi _{\rm b,u}=\frac{1}{2} \partial n_{\rm b,u}/\partial \mu_{\rm b,u}|_{\mu_{u,b}}$, 
where $\mu_{\rm b}=\mu +\epsilon_b/2$, $\mu_{\rm u}=\mu+H/2$ and the total density $n$ is fixed. 
Here $n_b$ and $n_u$ are the densities of pairs and excess fermions. 
Physically, the system has two processes occurring in parallel, namely the breaking of pairs and the alignment of spins.
The analog for the zero temperature susceptibility of the gapless phase is thus two parallel resistors in a circuit.

We also find that  the effective susceptibilities for the TLL of bound pairs and the TLL of excess fermions are expressed as 
$\chi_{\rm b}=1/(\hbar \pi v_N^{b})$ and  $\chi_{\rm u}=1/(4\hbar \pi v_N^{{\rm u}})$. 
The  density stiffness parameters are obtained from 
$v_N^r=\frac{L}{\pi\, \hbar }\frac{\partial ^2E_0^{r}}{\partial N_r^2}$ for a Galilean invariant system, 
with $r=1$ for excess fermions and $r=2$ for bound pairs. For the strongly interacting regime ($\gamma >1$), 
the  ground state energies for the  pairs and excess fermions are given explicitly by \cite{Guan2007} 
$E_0^{r}\approx\frac{\hbar^2}{2m}  \frac{\pi^2N^3}{3rL^2}\left(1+\frac{2A_r}{|c|}+\frac{3A^2_r}{c^2} \right)$ 
with $A_1=4n_2$ and $A_2=2n_1+n_2$.  
Here $n_1$ and $n_2$ are the density of excess fermions and pairs, respectively. 
Thus
\begin{eqnarray}
v_{N}^{{\rm b}}&=& \frac{\hbar \pi n_2}{2m}\left[ 1+\frac{4}{|c|}(n-3n_2)+\frac{3}{c^2}(4n^2-24nn_2+30n_2^2)\right]\nonumber\\
v_N^{{\rm u}}&=&\frac{\hbar\pi n_1 }{m}\left[ 1+\frac{4}{|c|}(n-2n_1)+\frac{4}{c^2}(3n^2+10n_1^2-12nn_1)   \right]. \nonumber
\end{eqnarray}

\begin{figure}[htbp]
\includegraphics[width=0.99\linewidth]{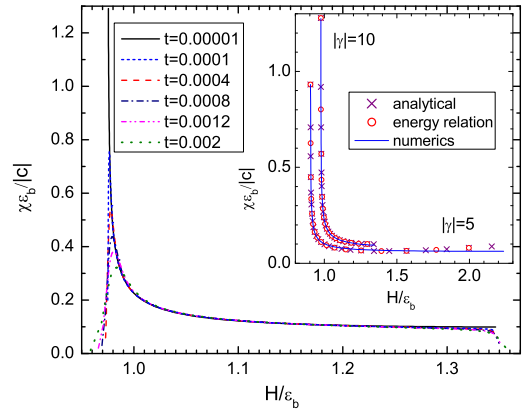}
\caption{ (Color online) The dimensionless susceptibility vs magnetic field for $|\gamma|=10$ at different temperatures.  
The susceptibility is independent of temperature for $T< H-H_{c1}$ and $T<H_{c2}-H$. Round peaks of the 
susceptibility in the vicinity of the two critical points are observed at low temperatures.
The inset shows the  susceptibility for $|\gamma| =5$ and $10$ at $T=0$. 
The pink  crosses denote the analytic result (\ref{susceptibility}) which is   in excellent agreement with   
the numerical  results obtained from the field-magnetization relation \cite{Guan2007} (red circles) 
and from the dressed energy equations  \cite{ref-sup}  (blue  lines).} \label{fig:sus-0}
\end{figure}

The analytic expression (\ref{susceptibility}) with these velocities is in excellent agreement with  
the numerical results (see inset in Fig.~\ref{fig:sus-0}).

The onset susceptibility at  the  lower and upper critical fields $H_{c1}$ and $H_{c2}$ is 
related to the collective nature of the pairs and excess fermions, with 
\begin{eqnarray}
 \chi \bigm|_{H\to H_{c1} +0}&=&\frac{1}{\hbar \pi v_N^{\rm b }}\biggm|_{n_2=\frac{n}{2}}=\frac{K^{\rm b}}{\hbar \pi v_s^{\rm b }} \biggm|_{n_2=\frac{n}{2}},\\
  \chi \bigm|_{H\to H_{c2} -0}&=&\frac{1}{4\hbar \pi v_N^{\rm u }}\biggm|_{n_1=n}=\frac{K^{\rm u}}{4\hbar \pi v_s^{\rm u }}\biggm|_{n_1=n}.
\end{eqnarray}
Here $v_s^r$ and $K^{r}=v_s^{r}/v_N^{r}$  are the sound velocities and effective TLL parameters of the 
bound pairs and excess single fermions. 
From the relation $v_s^{r} =\sqrt{\frac{L}{mn}\frac{\partial ^2E_0^{r}}{\partial L^2}}$,  the velocities are given by 
$v_s^{r}=\frac{\hbar }{2m}\frac{2\pi n_r }{r}  \left(1+ 2A_r/|c| +3A_{r}^2/c^2 \right)$.

%
The separation of the susceptibility (\ref{susceptibility}) naturally suggests that the  low energy physics of the polarized pairing phase is  described by a renormalization fixed point of the two-component TLL class, 
where the interaction effect enters into the collective velocities, or equivalently the effective masses of the two TLLs are varied by the interaction.
At finite low temperatures,  the two-component  TLL acquires a universal form 
$ F(T,H)\approx  E_0(H)-\frac{\pi   k_B^2T^2}{6\hbar}\left(1/v_s^{\rm b}+1/v_s^{\rm u}\right)$ of the free energy.
For temperature $T<H-H_{c1}$ and $T<H_{c2}-H$, the susceptibility is indeed independent of temperature 
provided that $-\partial^2 \left(1/v_s^{\rm b}+1/v_s^{\rm u}\right)/\partial H^2\approx 0 $, see Fig.~\ref{fig:sus-0}. 
We clearly see that the $T=0$ divergent susceptibility near the critical point $H_{c1}$ evolves into  round peaks at low   temperatures.  
The peak hight  decreases as the temperature increases. Here the leading irrelevant operators gives a 
correction of the order $O(T^2)$ to the low energy in the vicinities of the two critical points.

For the quantum critical regime ($T>H-H_{c1}$ and $T>H_{c2}-H$) the susceptibility defines the universality class 
for quantum criticality of nonrelativistic Fermi theory, with \cite{ref-sup}
\begin{equation}
\chi \sim \frac{|c|}{\epsilon_{\rm b}}\left[\lambda_0+
\lambda_s t^{\frac{d}{z}+1-\frac{2}{\nu z}} {\rm Li}_{-\frac{1}{2}}  
\left( -\mathrm{e}^{\frac{\alpha(h-h_{c1})}{t^{\frac{1}{\nu z}}}  } \right)\right] .  \label{QC-Hc}
\end{equation}
Near the critical point $h_{c1}=-2\tilde{\mu} +\frac{32}{3\pi\sqrt{2}}(\tilde{\mu} +1/2)^{3/2}$ we have $\lambda_0=0$ 
and $\lambda\approx \frac{1}{8\sqrt{2}\pi }\left(1-\frac{6}{\pi}\sqrt{(h-h_{c1})/2} \right)$ with $\alpha=1/2$, $t=T/\epsilon_b$ and $h=H/\epsilon_b$. 
Here the dynamical critical exponent $z=2$ and correlation length exponent $\nu =1/2$ for different phases of the spin states.   
Near the upper critical point $h_{c2}$ the susceptibility defines a similar form as (\ref{QC-Hc}), 
but with the background susceptibility $\lambda_0\ne0$ \cite{ref-sup}.

\textit{Specific heat.-} We turn now to the specific heat of the attractive Fermi gas. 
The low temperature expansion of the TBA equations with respect to $T\ll H,\epsilon_b $ gives
\begin{equation}
c_v=\frac{\pi  k_B^2 T}{3\hbar}\left(\frac{1}{v_s^{\rm b}}+\frac{1}{v_s^{\rm u}}\right). \label{heat}
\end{equation}
The linear $T$-dependence of the specific heat is a consequence of linear dispersions in branches of pairs and single fermions. 
The breakdown of this linear temperature-dependent relation defines a crossover temperature $T^*$ which charaterizes  
a universal crossover from a relativistic  dispersion into a  nonrelativistic  dispersion \cite{Maeda2007,Zhao}.

\begin{figure}[htbp]
\includegraphics[width=0.99\linewidth]{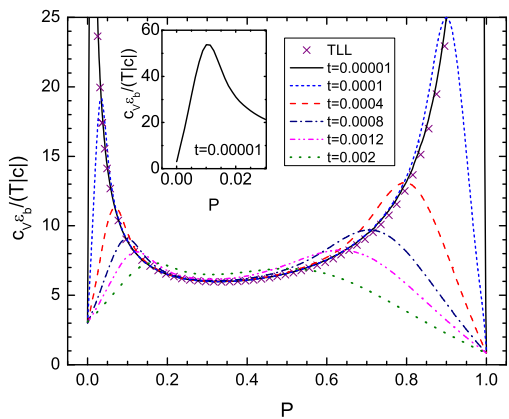}
\caption{ (Color online) Dimensionless specific heat  vs polarization for $|\gamma|=10$ at different temperatures. 
The deviation from linear temperature dependence (\ref{heat})  (red crosses) indicates the breakdown of the two-component TLL.  
The  inset shows a round peak evolved near $H_{c1}$ at $T=0.00001\epsilon_b$.} 
\label{fig:cv}
\end{figure}

We see clearly in Fig.~\ref{fig:cv} that at low temperatures a peak evolves in the  specific heat near 
each of the two critical points, i.e., near $P=0$ and $P=1$ due to a sudden change in the density of states.  
We also note that the peak positions  mark the TLL specific heat curve (\ref{heat}). 
The two peaks merge at  the top of the TLL phase in Fig.~\ref{fig:contour}.  
Thus the peak  position in turn gives the TLL phase boundary in the $c_v-P$ or $c_v-H$ plane.  
The specific heat obtained from the equation of state \cite{Guan-Ho} also defines a scaling behaviour 
\begin{equation}
c_v\sim \sqrt{ \frac{2m\varepsilon_b} {\hbar^2 t^2}} \left[\nu_0+\nu_s t^{\frac{d}{z}+1-\frac{2}{\nu z}} {\rm Li}_{-\frac{1}{2}}  
\left( -\mathrm{e}^{\frac{\alpha(h-h_{c})}{t^{\frac{1}{\nu z}}}  } \right)\right] \label{QC-cv}
\end{equation}
where $\nu_0$, $\nu_s$ and $\alpha $ are constants which can be determined from the closed form of the specific heat if necessary \cite{ref-sup}. 
The two-component TLL specific heat (\ref{heat}) is clearly manifest in Fig.~\ref{fig:cv} from the numerical result obtained using the equation of state. 

\begin{figure}
\includegraphics[width=0.99\linewidth]{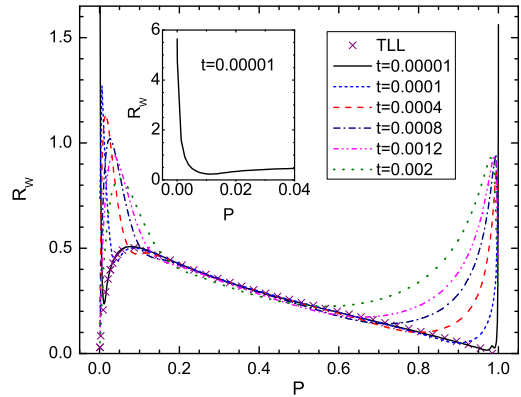}
\caption{ (Color online) Wilson Ration vs polarization  for  $|\gamma|=10$ at different temperatures.   
The numerical result obtained from the equation of state fully agrees with the Wilson ratio (\ref{W_R}) (red crosses)  
for the two-component TLL phase. The deviations from the  result (\ref{W_R}) characterise the crossover 
temperature $T^*$.  Anomalous behaviour is found near $P=0$ and $P=1$ 
(see inset for near the critical point $H_{c1}$). 
} \label{fig:Rw}
\end{figure}

\textit{Wilson ratio.-} The linear temperature-dependent nature of the specific heat and the 
separable feature of the susceptibility give the Wilson ratio (\ref{W_R}) 
for the effective low energy physics of the two-component TLL.
This Wilson ratio for the 1D attractive Fermi gas  is significantly different from the ratio obtained 
for the field-induced gapless phase in the quasi-1D gapped spin ladder \cite{Ninios2012}, 
where the gapless phase is a single-component TLL \cite{Schulz1991,WangYP} 
and the ratio gives a renormalization fixed point of a linear spin-$1/2$ chain in zero field. 
It is interesting to note that for the 1D attractive Fermi gas the onset Wilson ratio also depends solely on the TLL  parameters, with   
\begin{equation}
W_R \bigm|_{H\to H_{c1}}=4K^{\rm b} \bigm|_{n_2\to \frac{n}{2} },\quad W_R \bigm|_{H\to H_{c2}}= K^{\rm u} \bigm|_{n_1\to n}.\nonumber
\end{equation}
Here we find 
\begin{eqnarray}
K^{\rm b}&\approx & 1+\frac{6}{|c|}n_2 +\frac{3}{c^2}n_2(3n_2+4n)\nonumber\\
K^{\rm u}&\approx & 1+\frac{4}{|c|}n_1 +\frac{4}{c^2}n_1(n_1+2n).
\end{eqnarray}
Note that the values in the limit of infinitely strong coupling are $W_R=4$ at $H_{c1}$ and $W_R=1$ at $H_{c2}$.

The anomalous enhancement of the Wilson ratio near the onset values is shown in Fig.~\ref{fig:Rw}. 
Anomalous enhancement of the Wilson ratio has been observed near the metal-insulator transition in simulations of 
a three-dimensional quantum spin liquid \cite{3D}.
Here for the 1D attractive Fermi gases this anomalous divergence is mainly due to sudden changes in the density of states either in the bound state or excess fermion branch.  
Again, deviation from the Wilson ratio (\ref{W_R}) gives the crossover temperature $T^*\sim |H-H_{c}|$ separating   the TLL from   the free fermion liquid near the critical points.   
In addition to the anomalous divergence of the onset Wilson ratio, a round peak is observed near $P\approx 0.1$ 
due to the competing ordering of the two TLLs.  
$R_W < 1$ for  finite values of the polarization ($0<P<1$). 

In contrast to this enhancement,  for the repulsive regime the Wilson ratio is always less than $2$, i.e., $R_W=2/(1+v_{\sigma}/v_c)$ 
which  simply  gives a fixed point of  the TLL  in the context of  spin-charge separation. 
Here the charge and spin velocities $v_{c,\sigma}$ can be calculated following  \cite{LeePRB}.

The Wilson ratio of 1D Fermi gases can in principle be measured in experiments.
The finite temperature density profiles of a 1D trapped Fermi gas of $^6$Li atoms have been measured \cite{Liao}. 
Most recently, the susceptibility  has been directly obtained from the density profile of the trapped atomic cloud in higher dimensions \cite{Lee2013}. 
High precision measurements of thermodynamic quantities have also been reported \cite{Ku2012}.
For the 1D case, the predicted susceptibility could be tested from the density profiles $n_{\uparrow, \downarrow}$ and the chemical potential bias.

The Wilson ratio of the 1D attractive Fermi gases which we have obtained provides a measurable parameter to quantify different  phases of quantum liquids  
in 1D interacting fermions with polarization.  
At low temperatures, the Fermi liquid nature is retained in 1D many-body systems of interacting fermions. 
Our  analysis can be adapted to different systems, such as interacting fermions, bosons and mixtures composed of cold atoms with higher spin symmetry.

{\em Acknowledgments.} We thank M. Cazalila and W. Vincent Liu for helpful discussions.  
This work has been supported by the NNSFC under the grant No. 91230203 and the National Basic Research Program of China under Grants No.  2012CB922101
No. 2012CB821300,  and No. 2011CB922200.
The work of XWG and MTB  has been partially supported by the Australian Research Council. 
XWG thanks Chinese University of Hong Kong for kind hospitality. 
M. T. B. is supported by the 1000 Talents Program of China.
AF acknowledges financial support  from Coordena\c c\~ao de Aperfei\c coamento de Pessoal de Nivel Superior (Proc. 10126-12-0).


\onecolumngrid

\vskip 2cm
\centerline{\bf Supplementary material}
\vskip 0.5cm

The Gaudin-Yang model \cite{Yang,Gaudin} is exactly solved by means of the nested Bethe ansatz.  
The thermodynamics of the model is given explicitly in Takahashi's book  \cite{Takahashi}.
At finite temperatures, the density distribution functions of pairs, unpaired fermions and spin strings involve the densities of 
`particles' $\rho_r(k)$ and `holes' $\rho_r^h(k)$ ($r=1,2$ for single excess  fermions and bound pairs).  
Following the Yang-Yang grand canonical ensemble method,  the grand partition function 
$Z={\mathrm {tr}} (\mathrm{e}^{-\cal{H}/T})=\mathrm{e}^{-G/T}$ in terms of the Gibbs
free energy $G = E - HM^z - \mu n - TS$  with respect to the magnetic field $H$, 
chemical potential $\mu$ and entropy $S$.   
In terms of the dressed energies $\epsilon^{\rm b}(k) := T\ln( \rho_2^h(k)/\rho_2(k) )$ and 
$\epsilon^{\rm u}(k) := T\ln( \rho_1^h(k)/\rho_1 (k) )$ for paired and unpaired fermions,
the equilibrium states are determined by the minimization condition of
the Gibbs free energy, which gives rise to the set of coupled nonlinear
integral equations in terms of the dressed energies $\epsilon^{\rm b}$ and $\epsilon^{\rm u}$ 
\begin{eqnarray}
\epsilon^{\rm
  b}(k)&=&g^{\rm b}(k) +K_2*f_{\epsilon^{\rm b}}(k)+K_1*f_{\epsilon^{\rm u}} (k)\\
\epsilon^{\rm
  u}(k)&=&g^{\rm u}(k) +K_1*f_{\epsilon^{\rm b}}(k) -\sum_{\ell=1}^{\infty}K_\ell*f_{T\ln \eta_{\ell}}(k)\\
T\ln \eta_\ell (\lambda)&=&\ell H+K_\ell *f_{\epsilon^{\rm u}}(\lambda) +\sum_{n=1}^{\infty}T_{\ell m}*f_{T\ln \eta_m}(\lambda)
\end{eqnarray}
with  $\ell =1,\ldots, \infty$. 
The driving terms $g^{\rm b}(k)  = 2(k^2-\mu-{c^2}/4)$ and $g^{\rm u}(k) = k^2-\mu-{H}/{2}$. 
Here $*$ denotes the convolution integral
$K_m*f_x(\lambda) = \int_{-\infty}^\infty K_m (\lambda-\lambda') f_x(\lambda')  d\lambda'$ with 
$K_m(\lambda)=\frac{1}{2\pi}\frac{m|c|}{(mc/2)^2+\lambda ^2}$ and $f_x(k)=T\ln\left(1+\mathrm{e}^{-x(k)/T}\right)$.
The function $\eta_\ell (\lambda) = \xi^h_\ell(\lambda)/ \xi_\ell(\lambda)$ is the ratio of the string densities. 
The function $T_{\ell m}(k)$ is given explicitly by  \cite{Takahashi,Guan2007,Schlottmann2012}
 \begin{eqnarray}
T_{mn}(x)=\left\{
\begin{array}{ll}
    a_{|m-n|}(x)+2a_{|m-n|+2}(x)+ \ldots +2a_{m+n-2}(x)+a_{m+n}(x), & \quad\hbox{for $n\neq m$}
    \\
    2a_{2}(x)+2a_{4}(x) \ldots +2a_{2n-2}(x)+a_{2n}(x), & \quad\hbox{for $n=m$.} \\
\end{array}%
\right.\nonumber
\end{eqnarray}
The Gibbs free energy per unit length is given by $G=p^b+p^u$ where the effective pressures of the unpaired fermions and  bound pairs are given by
\begin{equation}
p^r=\frac{rT}{2\pi}\int_{-\infty}^{\infty}dk\ln(1+\mathrm{e}^{-\epsilon^{\rm r}(k)/{T}}) \nonumber
\end{equation}
with $r=1$ for unpaired fermions and $r=2$ for paired fermions.

The thermodynamics of the model can be calculated from the standard thermodynamic relations. 
The density, magnetization, entropy, susceptibility and specific heat  are given by 
\begin{eqnarray}
n&=&\left( \frac{\partial p(\mu,H,T) }{\partial \mu} \right)_{T,H},\qquad M^z=\left( \frac{\partial p(\mu,H,T) }{\partial H} \right)_{T,\mu},\nonumber\\
s &=& \left( \frac{\partial p(\mu,H,T) }{\partial T} \right)_{\mu,H},\qquad \chi= \left(\frac{\partial M^z}{\partial H} \right)_{n,T},\qquad c_v =T\left(\frac{\partial s}{\partial T} \right)_{n,p}.
\end{eqnarray}

The TBA equations provide the full thermodynamics of the model, including the Tomonaga-Luttinger liquid physics and quantum criticality. 
At zero temperature, the quantum phase diagram in the grand canonical ensemble can be analytically determined from the 
dressed energy equations \cite{Takahashi,Guan2007}
\begin{eqnarray}
\epsilon^{\rm b}(k)&=&g^b(k)-\int_{-Q_2}^{Q_2}K_2(k-\Lambda){\epsilon^{\rm   b}}(\Lambda)d\Lambda'
-\int_{-Q_1}^{Q_1}K_1(k-k'){\epsilon^{\rm u}}(k')d k'\nonumber\\
\epsilon^{\rm u}(k)&=&g^u(k)-\int_{-Q_2}^{Q_2}K_1(k-\Lambda){\epsilon^{\rm
    b}}(\Lambda)d\Lambda \nonumber
\end{eqnarray}
which are obtained from the TBA equations (11)-(13)  in the limit $T\to 0$. 
The dressed energy $\epsilon^{\rm b}(\Lambda)\le 0$
($\epsilon^{\rm u}(k)\le 0$) for $|\Lambda|\le Q_2$ ($|k|\le Q_1$) correspond
to the occupied states. The positive part of $\epsilon^{\rm b}$
($\epsilon^{\rm u}$) corresponds to the unoccupied states.
The integration boundaries $Q_{2}$ and $Q_1$  characterize the Fermi surfaces for
bound pairs and unpaired fermions, respectively. 
The pressures of pairs and excess fermions are given by 
\begin{equation}
 p^{\rm b}=- \frac{1}{\pi}\int_{-Q_2}^{Q_2} d \Lambda \, \epsilon^{\rm  b}(\Lambda),\quad p^{\rm u}= -\frac{1}{2\pi}\int_{-Q_1}^{Q_1} dk \, \epsilon^{\rm  u}(k). 
 \nonumber
 \end{equation}      
The zero temperature susceptibility is obtained from these pressures using the standard statistical physics relations. 

In terms of the dimensionless quantities   
$\tilde{\mu}:= \mu/\varepsilon_{b}$, $h:= H/\varepsilon_{b}$, $t:= T/\varepsilon_{b}$ and 
$\tilde{n}:= n/|c|=\gamma^{-1}$, where $\varepsilon_b = \frac{\hbar^{2} }{2m} \frac{c^2}{2}$ is the binding energy,  
the equation of states for the strongly attractive gas  is   \cite{Guan-Ho}
 \begin{equation}
 \tilde{p}(t, \tilde{\mu}, h) := p/(|c|\varepsilon_b)=\tilde{p}^b+\tilde{p}^{u}, \label{EOS}
 \end{equation}
where  the pressures of the bound pairs and unpaired fermions are given by
 \begin{eqnarray}
\tilde{p}^b&=&-\frac{t^{\frac{3}{2}}}{2\sqrt{\pi}}F_{3/2}^{b}\left[1+\frac{\tilde{p}^b}{8}+ 2\tilde{p}^u\right]+O(c^4) \nonumber\\
\tilde{p}^u&=&-\frac{t^{\frac{3}{2}}}{2\sqrt{2\pi}}F_{3/2}^{u}\left[1+ 2\tilde{p}^b\right]+O(c^4) \nonumber
\end{eqnarray}
in terms of the functions $F_n^b$, $F_n^u$, $f_n^b$, and $f_n^u$ defined by 
$ F_n^{b,u} := {\rm Li}_n\left(-{\rm e}^{{X_{b,u}}/{t}}\right)$ and $f_n^{b,u} := {\rm Li}_n \left(  -{\rm e}^{{\nu_{b,u}}/{t}}\right)$,
with the notation $\nu_{b} = 2{\tilde{\mu}}+1$, $\nu_{u} =  \tilde{\mu} + h/2$. 
Here ${\rm Li}_{s}(z) = \sum_{k=1}^{\infty}z^{k}/k^{s}$ is the polylog function, with $I_0(x)=\sum_{k=0}^{\infty}\frac{1}{(k!)^2}(\frac{x}{2})^{2k}$ and 
\begin{eqnarray}
\frac{X_b}{t}&=&\frac{\nu_b}{t}-\frac{\tilde{p}^b}{t}-\frac{4\tilde{p}^u}{t}-\frac{t^{\frac{3}{2}}}{\sqrt{\pi}}\left( \frac{1}{16}f_{5/2}^{b}+\sqrt{2}f_{5/2}^{u}\right)\label{Xb}  \nonumber\\
\frac{X_u}{t}&=&\frac{\nu_u}{t}-\frac{2\tilde{p}^b}{t}-\frac{t^{\frac{3}{2}}}{2\sqrt{\pi}}f_{5/2}^{b}+\mathrm{e}^{-{h}/{t}} \mathrm{e}^{-K}I_0(K). \nonumber
\end{eqnarray}

From  the equation of states (\ref{EOS}),  the susceptibility $\tilde{\chi}=\chi \varepsilon_b /|c|$   at finite temperatures is given by 
\begin{eqnarray}
\tilde{\chi} =-\frac{1}{8 \sqrt{2\pi }\Delta^3}\left\{\frac{1}{\sqrt{t}}F_{-\frac{1}{2}}^{u}\left[1+\frac{3\sqrt{t}}{2\sqrt{\pi
}}F_{1/2}^{A_b}+ \frac{2\sqrt{2}t}{\pi}F_{\frac{1}{2}}^{b}F_{\frac{1}{2}}^{u}\right] 
+\frac{2\sqrt{2}\sqrt{t}}{\pi}F_{-\frac{1}{2}}^{b} \left(F_{\frac{1}{2}}^{u}\right)^2\right\} \nonumber
\end{eqnarray}
where
\begin{equation}
\Delta=1-\frac{\sqrt{t}}{2\sqrt{\pi }}F_{\frac{1}{2}}^{b}-\frac{t\sqrt{2}}{\pi}F_{\frac{1}{2}}^{b}F_{\frac{1}{2}}^{u}+\frac{t^{\frac{3}{2}}}{16\sqrt{\pi}}F_{3/2}^{b}. \nonumber
\end{equation}
In the quantum critical regime, i.e., in  the vicinity of the critical point and for temperature $T>|H-H_{c}|$, the universal scaling form can be evaluated analytically, with 
\begin{equation}
\chi \sim \frac{|c|}{\epsilon_{\rm b}}\left[\lambda_0+\lambda_s t^{\frac{d}{z}+1-\frac{2}{\nu z}} {\rm Li}_{-\frac{1}{2}}  
\left( -\mathrm{e}^{\frac{\alpha(h-h_{c})}{t^{\frac{1}{\nu z}}}  } \right)\right] \nonumber
\end{equation}
as given in the text. 
Near the critical point $h_{c2} \approx 1+(3\pi)^{2/3}(\tilde{\mu} +1/2)^{3/2}-2(\tilde{\mu}+1/2)$, we find 
$\lambda_0\approx 1/( 8\sqrt{2}\pi \sqrt{\lambda_2^{u}})$, $\lambda_s\approx \lambda_2^u/(\pi^2\sqrt{\pi} )$, 
$\alpha \approx \frac{1}{\sqrt{2} \pi}\left(3\sqrt{2}\pi (2\tilde{\mu} +1) \right)^{1/3}$ 
and $\lambda_2^u\approx \left( 3\sqrt{2} \pi(2\tilde{\mu} +1)/8 \right)^{2/3} -16 \left(\tilde{\mu}+1/2\right)^{3/2}/(3\sqrt{2}\pi )$.

Morover, by iteration, the specific heat can be obtained from the equation of states (\ref{EOS}) as $c_v=c_v^b+c_v^u$ where 
\begin{eqnarray}
\frac{c_v^b}{|c|}&=&\frac{1}{\sqrt{\pi} }\left\{ -\frac{3}{8}\sqrt{t}F_{\frac{3}{2}}^b+\sqrt{t}F_{\frac{1}{2}}^b\left(\frac{\tilde{\nu}_b}{2t}+\frac{5}{8t}(4\tilde{p}^u+\tilde{p}^b) 
+\frac{\sqrt{2}\tilde{\nu}_u}{\sqrt{\pi t}}F_{\frac{1}{2}}^u+\frac{\tilde{\nu}_b}{2\sqrt{\pi t}}F_{\frac{1}{2}}^b\right) \right. \nonumber\\
&&\left.-\frac{1}{2\sqrt{t}}F_{-\frac{1}{2}}^b\left( \frac{\tilde{\nu}_b}{t}(4\tilde{p}^u+\tilde{p}^b) 
+\frac{\tilde{\nu}_b^2}{t} +\frac{2\sqrt{2}\tilde{\nu}_b \tilde{\nu}_u }{\sqrt{\pi t}} F_{\frac{1}{2}}^u 
+\frac{3\tilde{\nu}_b^2}{2\sqrt{\pi t}} F_{\frac{1}{2}}^b+\frac{\tilde{\nu}^2_b}{\sqrt{2\pi t}} F_{\frac{1}{2}}^u\right)  \right\} \nonumber\\
\frac{c_v^u}{|c|}&=&\frac{1}{\sqrt{\pi} }\left\{ -\frac{3}{8\sqrt{2} }\sqrt{t}F_{\frac{3}{2}}^u
+\frac{\sqrt{t}}{2\sqrt{2}}F_{\frac{1}{2}}^u\left(\frac{\tilde{\nu}_u}{t}
+\frac{5}{2t}\tilde{p}^b+\frac{2\tilde{\nu}_b}{\sqrt{\pi t}}F_{\frac{1}{2}}^b\right)\right. \nonumber\\
&& \left.-\frac{1}{2\sqrt{2}\sqrt{t}}F_{-\frac{1}{2}}^u\left( \frac{\tilde{\nu}_u^2}{t} 
+\frac{2\tilde{\nu}_u}{t} \tilde{p}^b+\frac{2\tilde{\nu}_b \tilde{\nu}_u }{\sqrt{\pi t}} F_{\frac{1}{2}}^b 
+\frac{2\tilde{\nu}^2_u}{\sqrt{2\pi t}} F_{\frac{1}{2}}^b\right)  \right\} \nonumber .
\end{eqnarray}
The scaling form of the specific heat  in the quantum critical regime, i.e., $T> |H-H_c|$, can be worked out from these closed form expressions in a straightforward way, 
with the result given in the text.

\begin{figure}
\includegraphics[width=0.59\linewidth]{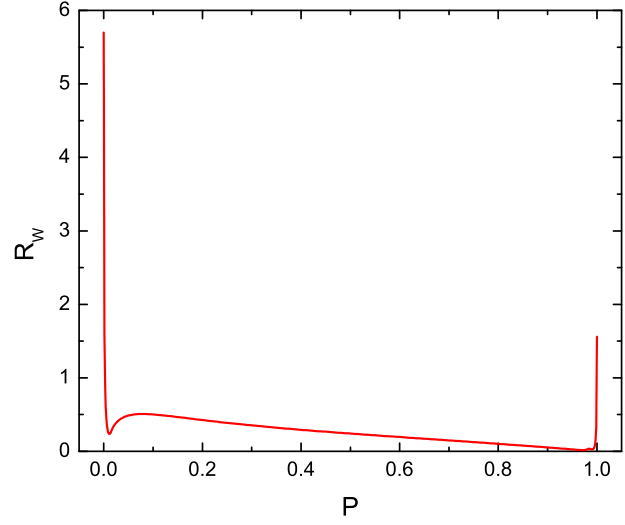}
\caption{ (Color online) Wilson ratio vs polarization for $|\gamma|=10$ at temperature $T=0.00001\epsilon_b$. 
The numerical result is obtained from the equation of states (\ref{EOS}). 
The ratio exhibits anomalous enhancement near the two critical points due to the sudden change of the density of states, 
where the values $R_W=5.53$ and $R_W=1.52$ agree with the values obtained from the analytic results (10). 
} 
\label{fig:Rw-Sup}
\end{figure}

The anomalous enhancement of the Wilson ratio exhibited near the two critical points is further demonstrated in Fig. \ref{fig:Rw-Sup}.

\end{document}